\documentclass[12pt]{article}
\usepackage{graphicx}
\usepackage{amsfonts}
\usepackage{latexsym}
\usepackage{amsmath}

\makeatletter
\@addtoreset{figure}{section}
\def\thefigure{\thesection.\@arabic\c@figure}
\@addtoreset{table}{section}
\def\thetable{\thesection.\@arabic\c@table}

\def\@sect#1#2#3#4#5#6[#7]#8{\ifnum #2>\c@secnumdepth
     \def\@svsec{}\else
     \refstepcounter{#1}\edef\@svsec{\csname the#1\endcsname.\hskip .75em
}\fi
     \@tempskipa #5\relax
      \ifdim \@tempskipa>\z@
        \begingroup #6\relax
          \@hangfrom{\hskip #3\relax\@svsec}{\interlinepenalty \@M #8\par}%
        \endgroup
       \csname #1mark\endcsname{#7}\addcontentsline
         {toc}{#1}{\ifnum #2>\c@secnumdepth \else
                      \protect\numberline{\csname the#1\endcsname}\fi
                    #7}\else
        \def\@svsechd{#6\hskip #3\@svsec #8\csname #1mark\endcsname
                      {#7}\addcontentsline
                           {toc}{#1}{\ifnum #2>\c@secnumdepth \else
                             \protect\numberline{\csname the#1\endcsname}\fi
                       #7}}\fi
     \@xsect{#5}}
\def\@begintheorem#1#2{\it \trivlist \item[\hskip \labelsep{\bf #1\ #2.}]}
\def\section{\@startsection {section}{1}{\z@}{-3.5ex plus -1ex minus
 -.2ex}{2.3ex plus .2ex}{\normalsize\bf}}

\begin{document}

\title{Why The Results of Parallel and Serial Monte Carlo Simulations May Differ}
\date{} 
\maketitle

\begin{center}
\author{
Boris D. Lubachevsky\\
{\em bdlubachevsky@aol.com}\\
}
\end{center}

\setlength{\baselineskip}{0.995\baselineskip}
\normalsize
\vspace{0.5\baselineskip}
\vspace{1.5\baselineskip}

\begin{abstract}
Parallel Monte Carlo simulations often expose faults in random number generators
\end{abstract}

A parallel Monte Carlo simulation is a sampling of a stochastic
process when this sampling is performed on concurrently active
multiple processors.
The counterpart serial Monte Carlo simulation
samples the same stochastic process but using a uniprocessor.
(Only the simulations in which
the process being sampled is supposed to be the same in both cases
are discussed here.)
Yet, in practice, it is often observed (but not as often reported)
that the stochastic properties of these two processes differ.

A typical example of such state of affairs is reported in \cite{MR03}, 
where a substantial difference between statistics obtained using 
a parallel algorithm introduced in \cite{L87} 
and the comparable statistics obtained using 
the corresponding serial algorithm
introduced in \cite{BKL75} is observed.
The authors in \cite{MR03} propose to compensate for the alleged damage
due to parallelization
(the true origin of which appears to be unknown to them, though some
speculations are offered)
with another damage that "bends the structure" in the opposite direction:
they modify the algorithm in \cite{L87} so as to fit the two outcomes.

A parallelization being done correctly, as in \cite{L87},
a mathematical IF-THEN theorem can be proven that assures the two stochastic
processes to be identical. Yet a computer experiment shows that the processes
differ. This may only mean the IF conditions of the theorem are not satisfied
in the experiment.  Where can the IF conditions fail?
The stochastic process generated by either serial or
parallel simulation is formed by feeding a source stochastic process,
based on a random number generator, to the deterministic mechanism of
the algorithm.  The theorem claims the two resulting processes,
for the parallel and for the serial algorithm, to be identical,
provided the source process satisfies certain properties.
That the resulting processes turn out to differ may only mean
that the source process does not satisfy the assumed properties.

The previous simple argument is general, applicable to many simulations.
For example, a difference in the simulation outcomes between a parallel
and the serial simulation was noticed in \cite{KLE96}. The simulation task
in \cite{KLE96} was rather different from that in \cite{MR03},
but the reason for the fault was the same: bad random number generator.

It is a well-known, textbook recommendation that a good random number generator
has to be employed if one expects to obtain statistically valid results
in Monte Carlo simulations.  A new "twist" is that
if the random number generator is not good, the faulty results will quite
probably be exposed during the parallel simulations but not necessarily
during the serial ones. The faults will be detected by comparing the
parallel runs with the serial runs or by comparing parallel runs among
themselves when those runs are made under different mappings of the task onto
the parallel machine and/or using different numbers of processors to host
the task.  By contrast, in only-serial Monte Carlo simulations,
there is usually no inherent mechanism to detect statistical errors.
Without obtaining comparable results in a different way,
such as via analytical estimates or by using differently arranged simulations,
the errors have a good chance to remain unnoticed.

For example, 
in \cite{MR03}, not only the reported statistics obtained in parallel
runs is incorrect, as it is noticed in \cite{MR03}, but the statistics
obtained in serial runs has to be faulty too, as long as the same
faulty random number generator is used. Yet, the authors in \cite{MR03} "bend"
only the parallel algorithm to eliminate the differences.
Apparently they trust the serial results.

Now, as we diagnosed the ailment, let us suggest a cure
that does not require one to "bend" good algorithms.
In the overwhelming majority of Monte Carlo simulations, the source stochastic
process mentioned above is a sequence of independent samples
of a random value uniformly distributed on the interval $(0,1)$.
The most likely culprit in most cases appears to be a
violation of the uniformity of the distribution.
That was established for the simulations
in \cite{KLE96} where the density of the distribution was found to
be smaller for smaller sampled values.

The way the errors appear in \cite{MR03}, a non-uniformity
seems to be also the reason for the discrepancy.
Specifically, a Poisson clock (parallel or serial) is
"ticking" in \cite{MR03} with increments proportional 
to $-log(x)$ where $x$ is sampled
on $(0,1)$.
If, like in \cite{KLE96}, 
smaller values of $x$ are less probable than larger values,
then the reported clock is slow and with each tick the average time lag of
the reported clock vs. the correct clock increases.
This would make the results of both parallel and serial runs inaccurate
but in different degrees because
parallel and serial runs differ in number of ticks
and in size of increments
for the same time increment of the correct clock.

A simple way to test the uniformity of the distribution of variable $x$
is to use another variable $y = f(x)$ instead of $x$ where $f$
transforms interval $(0,1)$ onto itself without changing uniformity.
For example, take $f(x) = f_1 (x) = 1-x$ or take $f(x) = f_2 (x) = x+1/2$
for $x < 1/2$ and $f_2 (x) = x-1/2$ for $x \ge 1/2$ or take a composition
$f(x) = f_1 (f_2 (x))$ and so on.
If statistical averages change, the distribution  of $x$ is
not uniform (and/or other faults are present in the random number generator).

A number of ways exist to fix the distribution non-uniformity.
A simple and practical fix is as follows.
Recognize a subinterval $(a,b)$, $a < b$,  of the interval $(0,1)$
such that the density of the distribution is satisfactorily uniform
on $(a,b)$. Instead of feeding in variable $x$, feed in
variable $y$ derived from $x$ as follows:
when the drawn $x$ does not belong to $(a,b)$,
discard that $x$ and draw again, and when
inequality $a < x < b$ holds, take $y = (x-a)/(b-a)$.


\begin{thebibliography}{MMMM}

\bibitem[MR03]{MR03} 
P.~A. Mahobar and A.~D. Rollett, Asynchronous Parallel Potts Model for
Simulation of Grain Growth,
Materials Science and Technology (MS and T '03)' conference
incorporating 'Modeling, Microstructure and Control in Ferrous and Non-Ferrous
Industry' symposium, ed. F. Kongoli et al., TMS and ISS, Chicago, Nov. 9 - 12,
pp. 399 - 412 (2003).
Also see: {\em mimp.mems.cmu.edu/papers/2003\_20.pdf}

\bibitem[L87]{L87} 
B.~D. Lubachevsky, Efficient Parallel Simulations of Asynchronous
Cellular Arrays, {\em Complex Systems}, {\bf 1} (1987), no. 6, 1099-1123,
S.Wolfram (ed.).  
Also see: {\em arXiv:cs/0502039}.

\bibitem[BKL75]{BKL75} 
A.~B. Bortz, et al., A New Algorithm for Monte Carlo
Simulation of Ising Spin Systems, {\em J. Comp. Physics}, {\bf 17} (1975), pp.10-18.
Also see: {\em tphex.hep.by/Journal\_of\_Computational\_Physics/Volume 17/1/2.pdf}

\bibitem[KLE96]{KLE96} 
K.~Kumaran et al., Massively Parallel Simulations of ATM Systems,
10th Workshop on Parallel and Distributed Simulations (PADS'96).
Also see: {\em www.computer.org/portal/web/csdl/doi/10.1109/PADS.1996.761561}

\end{thebibliography}
\end{document}